\documentclass[12pt,thmsa]{article}
\usepackage{amssymb}

\usepackage{sw20aip}

\input{tcilatex}
\begin{document}

\author{A. Miranda \\
Department of Physics and Astronomy\\
University of Aarhus, DK-8000\\
E-mail:miranda@phys.au.dk}
\title{Modelling the light quark vacuum}
\maketitle

\begin{abstract}
The triumphs of the Standard Model of Particle Physics call attention upon
an old idea, that the so-called vacuum is an accessible physical medium, and
not just a tautology.

I take this idea as a serious working hypothesis, and I suggest a new way of
trying to define and understand the structure of the light quark vacuum
within the general framework of the Standard Model.


BCS-like theory of quark-lepton vacuum
\end{abstract}

\section{Introduction}

We propose in this Letter that the physical vacuum, which has acquired a
place of honour in contemporary physics because of the great empirical
success of the Standard Model, is the ''ground state'' of a mostly unknown
physical medium that can be probed just like any other physical medium - and
therefore the idea is not just a tautology [1,2].

This idea has a long prehistory behind it. In modern physics, it first
emerged as a possibility, suggested by detailed QED studies. We need only
recall the Lamb shift, the anomalous magnetic moment of the electron, the
Casimir effect. This repertoire became greatly extended by the Standard
Model (the Higgs condensate, axial anomalies, radiative corrections of
various types, the so-called confinement of colour - just to name the most
famous).

Sections 2 and 3 introduce the Model Hamiltonian and Model Spaces assumed in
this paper. Here the guidelines for constructing these are provided by the
phenomenologically well-established Chiral Perturbation Theory of light
mesons and baryons and the various chiral quark models [3,4]. The
fundamental degrees of freedom are the chiral quarks of Perturbative QCD.
The gluons are not represented in this model space, as it is established
that there are no soft gluons. Neither are soft photons , as they are
assumed to be weakly coupled to quarks. Therefore the proposals in this
Letter are not immediately applicable to the lepton vacuum.

Sections 4 and 5 discuss the spectrum of u,d,s quasiparticles and the
stability of the vacuum defined to support such quasiparticles. The
empirical input is taken from the work of H.Leutwyler et al [5] .Particular
attention is paid to spinless and flavourless q\={q} vacuum stability modes.

Section 6 suggests possible experimental checks through electroweak probings
of the quark vacuum.

Finally, section 7 contains some concluding remarks.

\section{\protect\smallskip The Primary Model Hamiltonian H$^{(0)}$}

The model Hamiltonian, to be used in connection with a 3-momentum model
space $\mathcal{D}$ for QCD quarks, is assumed to have the particular form:

\begin{equation}
H^{(0)}[\mathcal{D]}=U_{B}^{(0)}+H_{0}^{(0)}+V^{(0)}[\mathcal{D}]
\end{equation}

If $\Omega $ represents the quantization box volume, then $%
U_{B}^{(0)}/\Omega $ is a ''background energy density '' as $\Omega
\rightarrow \infty $, supposed to absorb the infinite zero-point energy of
the virtual Dirac-Weyl quark seas. It is to be ''trivially renormalized ''
away.

\begin{equation}
H_{0}^{(0)}=\sum_{n}\sum_{\lambda }\int d^{3}\vec{p}|\vec{p}|(\alpha _{n\vec{%
p}\lambda }^{+}\alpha _{n\vec{p}\lambda }+\beta _{\bar{n}p\lambda }^{+}\beta
_{\bar{n}\vec{p}\lambda })
\end{equation}
\[
V^{(0)}[\mathcal{D}]=\sum_{n^{\prime },n}\sum_{\lambda ^{\prime }\lambda
}\int d^{3}\vec{p}^{\prime }\int d^{3}\vec{p}^{\prime }<n^{\prime }\lambda
^{\prime },\bar{n}^{\prime }\lambda ^{\prime };\vec{p}^{\prime }|V|n\lambda ,%
\bar{n}\lambda ;\vec{p}>\alpha _{n^{\prime }\vec{p}^{\prime }\lambda
^{\prime }}^{+}\beta _{\bar{n}^{\prime }-\vec{p}^{\prime }\lambda ^{\prime
}}^{+}\beta _{\bar{n}-\vec{p}\lambda }\alpha _{n\vec{p}\lambda } 
\]

\smallskip We assume that both $H_{0}^{(0)}$ and $V^{(0)}$ are hermitean.

The basic degrees of freedom defining our model space are Weyl fields, or
rather their Fourier components.

$n(\equiv u,d,s$ ) stands for particle species with momentum $\vec{p}$ and
helicity $\lambda $ ; $\bar{n}(\equiv \bar{u},\bar{d},\bar{s}$ ) represent
antiparticles, here defined to mean CP-conjugate states of quark states .

It is important to emphasize that interactions among the input Weyl fermions
are defined only within a \textit{momentum} space $\mathcal{D}$ extending
from zero up to about $\Lambda _{\chi }=1GeV$ (Section 5).This cut-off has
therefore \textit{a physical significance :} it cannot be removed without
removing the model space itself!

Colour degrees of freedom are not \textit{dynamically} involved in this
model . Their role is indirect, as they basically determine the \textit{%
choice }of the model Hamiltonian and the model spaces.Therefore they are not
explicitly represented.

\smallskip $\alpha _{n\vec{p}\lambda }^{+}(\alpha _{n\vec{p}\lambda })$ are
Fourier components of Weyl fields creating (annihilating) a quark $n\vec{p}%
\lambda $.

$\beta _{\bar{n}p\lambda }^{+}(\beta _{\bar{n}\vec{p}\lambda })$ are the
corresponding entities for their antiquarks.

As we shall see, the general idea is to introduce a series of canonical
transformations, starting from the fundamental Weyl fields, designed to
successively diagonalize as much of H$^{(0)}$ as possible. Small and
neglected residual contributions thus defined could then (hopefully !) be
added perturbatively, if necessary.

We begin by first choosing a Bogoliubov- Dirac canonical transformation
(without loss of generality for the purposes of this paper [5]):

\begin{equation}
\left( 
\begin{array}{l}
b_{n\vec{p}\lambda } \\ 
d_{\bar{n}-\vec{p}\lambda }^{+}
\end{array}
\right) =\left( 
\begin{array}{ll}
u_{np} & v_{np} \\ 
-v_{np}^{*} & u_{np}^{*}
\end{array}
\right) \left( 
\begin{array}{l}
\alpha _{n\vec{p}\lambda } \\ 
\beta _{\bar{n}-\vec{p}\lambda }^{+}
\end{array}
\right)
\end{equation}

\[
u_{np}=u_{\bar{n}p}\qquad v_{np}=v_{\bar{n}p} 
\]

The new operator set $b_{n\vec{p}\lambda }^{+},d_{\bar{n}\vec{p}\lambda
}^{+} $ shall be referred to as ''Dirac-Bogoliubov quasiparticle creation
operators'' or simply ''DB-quasiparticles''. Similarly for their
annihilation counterparts.We define the ''no particle state'' [5] for these
operators \textit{at time t}:

\begin{equation}
b_{n\vec{p}\lambda }(t)|0>_{t}=0
\end{equation}

\begin{equation}
d_{\bar{n}\vec{p}\lambda }(t)|0>_{t}=0
\end{equation}

\smallskip The parameters $u_{np},v_{np}$ play the role of variational
parameters at our disposal .

The folowing general condition upon the transformation matrix is imposed,
ensuring its invertibility:

\begin{equation}
|u_{np}|^{2}+|v_{np}|^{2}=1
\end{equation}

It is furthermore assumed (without loss of generality) that these
(variational) parameters are real.

\subsection{Fermionic Modes}

Given the Hamiltonian (1,2), the idea [5] is to solve the Heisenberg
equations of motion for constants of motion \textbf{\^{K}}(t):

\begin{equation}
i\frac{\partial }{\partial t}\mathbf{\hat{K}}(t)=[H^{(0)},\mathbf{\hat{K}}%
(t)]
\end{equation}

We begin by adjusting our variational parameters so that the quasiparticles
generated by the Dirac-Bogoliubov tranformation (3) \textit{are} \textit{%
stationary solutions }to the equations of motion (7) :

\begin{equation}
\lbrack H^{(0)},b_{\bar{n}\vec{p}\lambda }^{+}]=E_{np}b_{n\vec{p}\lambda
}^{+}+...
\end{equation}

\begin{equation}
\lbrack H^{(0)},d_{\bar{n}\vec{p}\lambda }^{+}]=E_{np}d_{\bar{n}\vec{p}%
\lambda }^{+}+...
\end{equation}

When computing the commutators, one finds both linear and non-linear terms
of course, but \textit{one extracts from the latter only linear
contributions, }if any\textit{,} and (at this stage) neglect\textit{\ }the
non-linear residuals.

Introducing the definitions

\begin{equation}
\mu \leftrightarrow (n\vec{p}\lambda )\ or\ (n\vec{p}\lambda ,\bar{n}-\vec{p}%
\lambda )
\end{equation}

\begin{equation}
\Delta _{\mu }\equiv -\sum_{\nu \in \mathcal{D}}V_{\mu \nu }u_{\nu }v_{\nu
}\equiv \Delta _{\mu }^{*}
\end{equation}

\begin{equation}
F_{\mu }\equiv V_{\mu \mu }v_{\mu }^{2}\equiv F_{\bar{\mu}}
\end{equation}

\begin{equation}
e_{\mu }\equiv \varepsilon _{\mu }+F_{\mu }\ 
\end{equation}

we find that

\begin{equation}
E_{\mu }^{2}=e_{\mu }^{2}+\Delta _{\mu }^{2}
\end{equation}

\begin{equation}
v_{\mu }^{2}=\frac{1}{2}(1-\frac{e_{\mu }}{E_{\mu }})
\end{equation}

\begin{equation}
u_{\mu }^{2}=\frac{1}{2}(1+\frac{e_{\mu }}{E_{\mu }})
\end{equation}

This fixes the variational parameters .

The function $e_{\mu }$ shall be referred to as ''the corrected single
particle energy'' , for obvious reasons.

The function $\Delta _{\mu }$ shall be referred to as ''the gap function'' ,
for reasons that should become clear in the sequel.

The function $E_{\mu }$ shall be referred to as ''DB-quasiparticle energy''.

Self-consistency requires that all gaps be either positive or
vanish.Solutions are discussed in Section 5.

\section{Model Hamiltonian H$^{(1)}$}

We shall continue approximating by next considering another branch of vacuum
excitations, more precisely, flavourless boson-like \textit{vacuum
excitation modes.} This should give some feeling about the \textit{dynamical 
}stability of the above vaccum solutions of the equations of motion, at
least in these most important channels .

Work continues within the framework of linearization procedures.

\smallskip Again, we seek \textit{stationary solutions} of the Heisenberg
equation of motion :

\begin{equation}
i\frac{\partial }{\partial t}\mathbf{\hat{K}}(t)=[H^{(1)},\mathbf{\hat{K}}%
(t)]
\end{equation}

We find

\[
H^{(1)}=const+\sum_{\mu }E_{\mu }(b_{\mu }^{\dagger }b_{\mu }+d_{\bar{\mu}%
}^{+}d_{\bar{\mu}})+\{\sum_{\mu ,\nu }V_{\mu \nu }u_{\mu }^{2}v_{\nu
}^{2}*b_{\mu }^{+}d_{\bar{\nu}}^{+}d_{\bar{\nu}}b_{\mu }+ 
\]

\[
+\sum_{\mu ,\nu }V_{\mu \nu }u_{\nu }v_{\nu }v_{\mu }u_{\mu }*b_{\nu }^{+}d_{%
\bar{\mu}}^{+}d_{\bar{\mu}}b_{\nu }+\sum_{\mu ,\nu }V_{\mu \nu }(u_{\mu
}^{2}u_{\nu }^{2}+v_{\mu }^{2}v_{\nu }^{2})*\Gamma _{\mu }^{+}\Gamma _{\nu } 
\]

\[
-\sum_{\nu ,\bar{\nu}^{\prime }}V_{\mu \nu }u_{\mu }^{2}v_{\nu }^{2}*(\Gamma
_{\mu }^{+}\Gamma _{\nu }^{+}+\Gamma _{\nu }\Gamma _{\nu })\}+\sum_{\mu ,\nu
}g_{\mu \nu }\Gamma _{\nu }^{+}(b_{\mu }^{+}b_{\mu }+d_{\bar{\mu}}^{+}d_{%
\bar{\mu}})+h.c. 
\]

where

\[
\Gamma _{\rho }^{+}\equiv b_{\rho }^{+}d_{\bar{\rho}}^{+} 
\]

\[
g_{\mu \nu }\equiv V_{\mu \nu }u_{\mu }v_{\mu }(u_{\nu }^{2}-v_{\nu }^{2}) 
\]

Once again we make a canonical transformation. Returning to the more
detailed notation, we introduce

\begin{equation}
\left( 
\begin{array}{l}
B_{f}(\vec{k}) \\ 
B_{f}^{+}(\vec{k})
\end{array}
\right) =\sum_{n\lambda }\int d^{3}\vec{p}\left( 
\begin{array}{ll}
X_{fn}(k,p) & -Y_{fn}(k,p) \\ 
-Y_{fn}k,p) & \ X_{fn}(k,p)
\end{array}
\right) \left( 
\begin{array}{l}
\Gamma _{n\lambda }(\vec{p}) \\ 
\Gamma _{n\lambda }^{+}(\vec{p})
\end{array}
\right)
\end{equation}

\begin{equation}
X_{fn}^{2}(k,p)-Y_{fn}^{2}(k,p)=1
\end{equation}

\begin{equation}
\lbrack H^{(2)},B_{f}^{+}(\vec{k})]=\omega _{f}(k)B_{f}^{+}(\vec{k})
\end{equation}

\smallskip

with the definitions

\begin{equation}
\Gamma _{n\lambda }^{+}(\vec{p})\equiv b_{n\vec{p}\lambda }^{+}d_{\bar{n}-%
\vec{p}\lambda }^{+}
\end{equation}

\begin{equation}
\omega _{f}(k)=\sqrt{k^{2}+\omega _{f}(0)}
\end{equation}

\smallskip The next effective Hamiltonian $\hat{H}^{(2)}[\mathcal{D}]$ is by
definition the Hamiltonian that makes equations (20) exact.Solutions must
satisfy condition (19). We illustrate the structure of the solutions in case
of complete degeneracy $m_{u}=m_{d}=m_{s}$.

\section{On electroweak probings of the vacuum}

The most straigtforward way of confronting these (or similar) \textit{%
ansatzes} on the quark vacuum with more or less direct experimental tests is
to make use of the hitherto ignored couplings to the electroweak sector of
the Standard Model [2,3].

We assume that, within the confines of Model Spaces chosen in this paper,
these \textit{model }couplings are given by the Model Hamiltonian

\begin{equation}
H^{(4)}[\mathcal{D}^{\prime }]=H^{(3)}[\mathcal{D}]+V^{(3)}+H_{0}+V_{ew}
\end{equation}

where $H_{0}$ describes free leptons and photons. ''Residual'' couplings are
(2,3)

\begin{equation}
V^{(3)}=\sum_{\mu ,\nu }g_{\mu \nu }\Gamma _{\nu }^{+}(b_{\mu }^{+}b_{\mu
}+d_{\bar{\mu}}^{+}d_{\bar{\mu}})+h.c.
\end{equation}

\begin{equation}
V_{ew}=e_{0}\int d^{3}\vec{r}J_{\alpha }^{em}(\vec{r},t)A^{\alpha }(\vec{r}%
,t)+\frac{G_{V}}{\sqrt{2}}\int d^{3}\vec{r}J_{\alpha }^{weak}(\vec{r}%
,t)l^{\alpha }(\vec{r},t)+h.c.
\end{equation}

The model space is now extended to $\mathcal{D}^{\prime }$ which includes in
addition leptons and photons.

\smallskip We make the usual ansatz for mesons (2,3):

\begin{equation}
|M;\vec{p}>\equiv B_{M\vec{p}}^{+}|0>=\sum_{\nu _{1}\oplus \bar{\nu}%
_{2}=M}\int d^{3}\vec{p}_{1}\int d^{3}\vec{p}_{2}
\end{equation}

\[
\Psi _{M\nu _{1}\bar{\nu}_{2}}(\vec{p}_{1},\vec{p}_{2})\delta (\vec{p}_{1}+%
\vec{p}_{2}-\vec{p})b_{\nu _{1}\vec{p}_{1}}^{+}d_{\bar{\nu}_{2}\vec{p}%
_{2}}^{+}|0> 
\]

with appropriate normalization condition on the wavefunctions. Amplitudes
for various electroweak processes can now be computed perturbatively [5].

\section{Parameters}

The extensive analysis carried out by Leutwyler et al [6] revealed that the
light quark obey the condition that their running masses in the $\stackrel{%
---}{MS}$ scheme at scale $\mu =1GeV$ must be

\begin{equation}
m_{u}=(5.1\pm 0.9)\ MeV\qquad m_{d}=(9.3\pm 1.4)MeV\qquad
\end{equation}

\[
m_{s}=(175\pm 25)MeV 
\]

We use this information to fix our basic parameters.

Using the effective interaction

\begin{equation}
<n^{\prime }\lambda ^{\prime },\bar{n}^{\prime }\lambda ^{\prime };\vec{p}%
^{\prime }|V|n\lambda ,\bar{n}\lambda ;\vec{p}>=-\frac{G_{n^{\prime }n}}{%
8\pi \Lambda _{\chi }^{2}}
\end{equation}

\medskip we find that the gap equations can be rewritten as

\begin{equation}
1=G_{1}I_{u}+G_{2}\varkappa _{1}I_{d}+G_{3}\varkappa _{2}I_{s}
\end{equation}

\begin{equation}
1=G_{2}\frac{I_{u}}{\varkappa _{1}}+G_{1}I_{d}+G_{3}\frac{\varkappa _{2}}{%
\varkappa _{1}}I_{s}
\end{equation}

\begin{equation}
1=G_{3}[\frac{I_{u}}{\varkappa _{2}}+\frac{\varkappa _{1}}{\varkappa _{2}}%
I_{d}+I_{s}]
\end{equation}

with the definitions

\begin{equation}
I_{k}=\Delta _{k}^{2}f(1/\Delta _{k})\qquad k=u,d,s
\end{equation}

\begin{equation}
f(u)=-\frac{1}{2}\ln [u+\sqrt{1+u^{2}}]+\frac{1}{2}u\sqrt{1+u^{2}}
\end{equation}

We are thus simply trading the input (27) with the G's, so that these
numerically fix \textit{all} our subsequent results and predictions.

We shall implement this kind of renormalization by following the example of
a single gap equation:

\[
\Delta =G*\int_{0}^{1}dxx^{2}u(x)v(x)=\Delta *\frac{G}{2}\int_{0}^{1}dx\frac{%
x^{2}}{\sqrt{x^{2}+\Delta ^{2}}} 
\]

or 
\begin{equation}
1=\frac{G\Delta ^{2}}{2}*f(1/\Delta )
\end{equation}

Lowering the upper integration limit from 1 downwards to some value $x$ we
find that this gap equation is satisfied if the parameter G is properly
adjusted:

\begin{equation}
1=\frac{g(x)}{2}*\Delta ^{2}f(x/\Delta )\qquad G=g(1)\quad
\end{equation}

\smallskip The results are (u in units of 1Mev):

\[
\begin{tabular}{cccc}
u & G$_{1}$ & G$_{2}$ & G$_{3}$ \\ 
10 & -0.3618 & 0.0167 & 0.2232 \\ 
20 & -0.0959 & 0.0030 & 0.0349 \\ 
30 & -0.0468 & 0.0012 & 0.0119
\end{tabular}
\]

The question of the dynamical stability of these solutions shall be
illustrated qualitatively. We thus ignore that the quasiparticles have in
fact different gap parameters and set these to a common value:

\begin{equation}
\Delta _{u}=\Delta _{d}=\Delta _{s}=\Delta
\end{equation}

\begin{equation}
G_{1}=G_{2}=G_{3}=\bar{G}/3
\end{equation}

The answer boils down to showing that the determinantal equation

\begin{equation}
S(\omega )=\left| 
\begin{array}{ll}
S_{11}(\omega ) & S_{12}(\omega ) \\ 
S_{21}(\omega ) & S_{22}(\omega )
\end{array}
\right| =0
\end{equation}

where

\begin{equation}
S_{11}(\omega )=1-\bar{G}(x_{\max })\mathbf{P}\int_{0}^{x_{\max }}duu^{2}%
\frac{2E(u)}{(2E(u))^{2}-\omega ^{2}(k)}
\end{equation}

\begin{equation}
S_{12}(\omega )=-\bar{G}(x_{\max })\omega (k)\mathbf{P}\int_{0}^{x_{\max
}}duu^{2}\frac{1}{(2E(u))^{2}-\omega ^{2}(k)}
\end{equation}

\begin{equation}
S_{21}(\omega )=S_{12}(\omega )
\end{equation}

\begin{equation}
S_{22}(\omega )=1-\bar{G}(x_{\max })\mathbf{P}\int_{0}^{x_{\max }}duu^{2}%
\frac{2E(u)\alpha ^{2}(u)}{(2E(u))^{2}-\omega ^{2}(k)}
\end{equation}

with

\begin{equation}
\alpha (u)=\sqrt{1-(\frac{\Delta }{E(u)})^{2}}
\end{equation}

has non-trivial solutions with $\omega \geq 0$.

Self energy corrections to DB-quasiparticles are not included. So $\omega =0$
is a solution, as long as the gap equation is satisfied. This is the
Goldstone associated with the quantum mechanical breaking of the symmetry
symbolized by conditions (39). The inclusion of self-energies would move
this Goldstone away from zero.Ignoring self-energy corrections, we estimate
that the non-Goldstone heavy boson has a rest mass (in units of 1GeV)
roughly of the order

\[
\omega ^{2}(0)=4\Delta ^{2}\{1+[\frac{1}{\Delta }\ \frac{\sqrt{1+\Delta ^{2}}%
-\Delta }{\ln \frac{1+\sqrt{1+\Delta ^{2}}}{\Delta }}]^{2}\} 
\]

\section{CONCLUDING REMARKS\textbf{\ }}

The physics of chiral and flavour symmetries and symmetry breaking in the
world of quarks and leptons, as assumed and described by the Standard Model
[2], gives us a deeper ''feeling'' for the unknown medium to which one
implicitly refers.

This Letter is inspired by the Chiral Perturbation Theory and the Chiral
Quark Model (2,3) and asks how one could interpret their empirical successes
in terms of\textit{\ } quantum motions of underlying \textit{QCD-quark
fields in their (suitably defined) vacuum state, }at least for mesons that
are apparently\textit{\ strongly entangled with the quark vacuum. }

A model Hamiltonian (and associated model spaces) are chosen to contain 
\textit{one} \textit{kind of degrees of freedom} \textit{only}, viz. the
so-called \textit{current quarks}, which we take to be spin-$\frac{1}{2}$
Weyl fermions. This input largely determines the choice of the vacuum 
\textit{ansatz} and the appropriate Model Hamiltonian.

The weight in this article is put on quantum dynamics and not on symmetries.
Thus isospin symmetry in particular is what it seems to be,i.e. an \textit{%
hadronic }symmetry.

The simplest way to further test and develop these ideas is through
electroweak probings (section 6).

\smallskip

\textbf{REFERENCES }

[1]T.D.Lee, Nucl.Phys.\textbf{A}538(1992)3c-14c.

[2] G. E.Volovik, The Universe in a Helium Droplet (OXFORD University Press,
2003).

[3] H. Georgi, Weak Interactions and Modern Particle Theory (Addison-Wesley
Publishing Company,1984).

[4] L. Glozman, A reply to Isgur's critique, nucl-th/9909021

[5]\nolinebreak \textbf{\ }P.A.M. Dirac\textbf{, }a) Lectures on Quantum
Field Theory, Academic Press, Inc.,New York,1966);b) in Nature \textbf{4941}%
,115(1964); c) Phys. Rev. \textbf{B139}, 684 (1965); d) Physics To-day 
\textbf{23,}29(1970).

[6] H. Leutwyler, Light Quark Masses,arXiv: hep-ph/9609467v1, 25th September
1996.

\smallskip

\end{document}